\documentclass{iau} 
\usepackage{graphicx}

\title[Swift J1644+5734: the EVN view] 
{Swift J1644+5734: \\ the EVN view}

 \author[Z. Paragi et al.]   
 {Z. Paragi$^1$,
 J. Yang$^{1,2,3}$,
S. Komossa$^4$,
 A. van der Horst$^5$,
 L.\,I.~Gurvits$^{1,6}$,
 R.\,M. Campbell$^1$,
 D. Giannios$^7$,
 \and T. An $^{3,8}$
 }

\affiliation{$^1$Joint Institute for VLBI ERIC (JIVE), Postbus 2, NL-7990 AA Dwingeloo, the Netherlands
\\ email: {\tt zparagi@jive.eu} \\[\affilskip]
$^2$ Dept. of Earth and Space Sciences, Chalmers Univ. of Technology, Onsala Space Observatory
\\ SE-439 92 Onsala, Sweden \\[\affilskip]
$^3$ Shanghai Astronomical Observatory, Chinese Acad. of Sciences, 200030 Shanghai, P.R. China \\[\affilskip]
$^4$ QianNan Normal University for Nationalities, Longshan Street, Duyun City of Guizhou Province, China \\[\affilskip]
$^5$ Dept. of Physics, The George Washington University, 725 21st Street NW, Washington, DC 20052, USA \\[\affilskip]
$^6$ Dept. of Astrodynamics and Space Missions, Delft University of Technology, NL-2629 HS Delft, the Netherlands \\[\affilskip]
$^7$ Dept. of Physics and Astronomy, Purdue University, 525 Northwestern Avenue, West Lafayette, IN 47907, USA \\[\affilskip]
$^8$ Key Lab. of Radio Astronomy, Chinese Academy of Sciences, 210008 Nanjing, P.R. China \\[\affilskip]
}

\pubyear{2017}
\volume{324}  
\jname{New Frontiers in Black Hole Astrophysics}
\editors{A.C. Editor, B.D. Editor \& C.E. Editor, eds.}
\begin{document}

\maketitle

\begin{abstract}
A small fraction of Tidal Disruption Events (TDE) produce relativistic jets, evidenced 
by their non-thermal X-ray spectra and transient radio emission. Here we present
milliarcsecond-resolution imaging results on TDE J1644+5734 
with the European VLBI Network (EVN). These
provide a strong astrometric constraint  on the average apparent  jet velocity 
$\beta_{\rm app}<0.27$, that constrains the intrinsic jet velocity
for a given viewing angle. 

\end{abstract}

\firstsection 
\section{Introduction}

Tidal Disruption Events (TDE) may help us understand
two outstanding questions of black hole astrophysics: how black holes form and grow, and 
what impact they have on the evolution of their host galaxies (see a recent review by 
\cite[Komossa 2015]{K15}).  The first question can be addressed through constraining the 
low-mass end of the supermassive black hole mass-function (\cite[Stone \& Metzger 2016]{SM16}). 
The idea here is that TDEs \lq\lq light-up'' dormant black holes (\cite[Giannios \& Metzger 2011]{GM11})
that have not shown an active galactic nucleus (AGN) signature before. As the size of both the
black hole and the disrupted star can be estimated well (e.g. \cite[Gezari et al. 2012]{Ge12}, 
\cite[Guillochon et al. 2014]{Gu14}), 
TDEs provide means to weigh black holes down to $10^4-10^5 M_{\odot}$, the majority of which
has remained undetected to date. The second question can be addressed through observations
of newly formed radio jets (triggered by a TDE) on milliarcsecond scales. Very long baseline
interferometry (VLBI) is a powerful tool to localise the origin of H\,{\sc I} outflows and to reveal the 
dynamics of atomic gas swept-up by the radio jet, catching AGN jet feedback in action 
(\cite[Morganti et al. 2013]{M13}); while VLBI of jetted TDEs provide us with a unique new chance 
of observing and understanding the very early phases of jet formation and evolution in an 
otherwise pristine environment, without any past AGN activity.

Future blind radio surveys will potentially detect a large number of TDEs, but it has been recognised
that a multi-band approach is essential for their identification. The role of VLBI may be two-fold: 
very high resolution observations of strong candidates may reveal their nuclear origin, supporting
the TDE identification at early times. On the other hand, securely identified TDEs (from X-ray properties
and/or optical spectra) that occur off-nucleus may point to peculiar systems such as a white dwarf
disrupted by an intermediate-mass black hole (\cite[Shcherbakov et al. 2013]{SHC13}), a TDE from 
a recoiling black hole (\cite[Komossa \& Merritt 2008]{KM08}), or it may reveal a secondary galactic 
nucleus in a merger (\cite[Liu \& Chen 2013]{LC13}). The latter is potentially interesting since the 
TDE rate is apparently higher in a certain type of post-starburst galaxies that have recently 
undergone a major merger event (\cite[French, Arcavi \& Zabludoff 2016]{FEZ16}).
It has been shown that jetted TDEs will be more efficiently detected in large-scale radio surveys than
with current or near-future X-ray missions, therefore TDEs will be primary targets of synchrotron  
transients with the Square Kilometre Array (SKA); the expected numbers however depend strongly on 
the fraction of jetted events, and their relativistic beaming factors (\cite[Donnarumma \& Rossi 2015]{DR15}). 
For the first time, we obtained a strong constraint on the average apparent jet speed for J1644+5734, 
briefly described below. More details of this result can be found in \cite{Y16}.

\begin{figure}[t]
\begin{center}
  \includegraphics[width=4.1in]{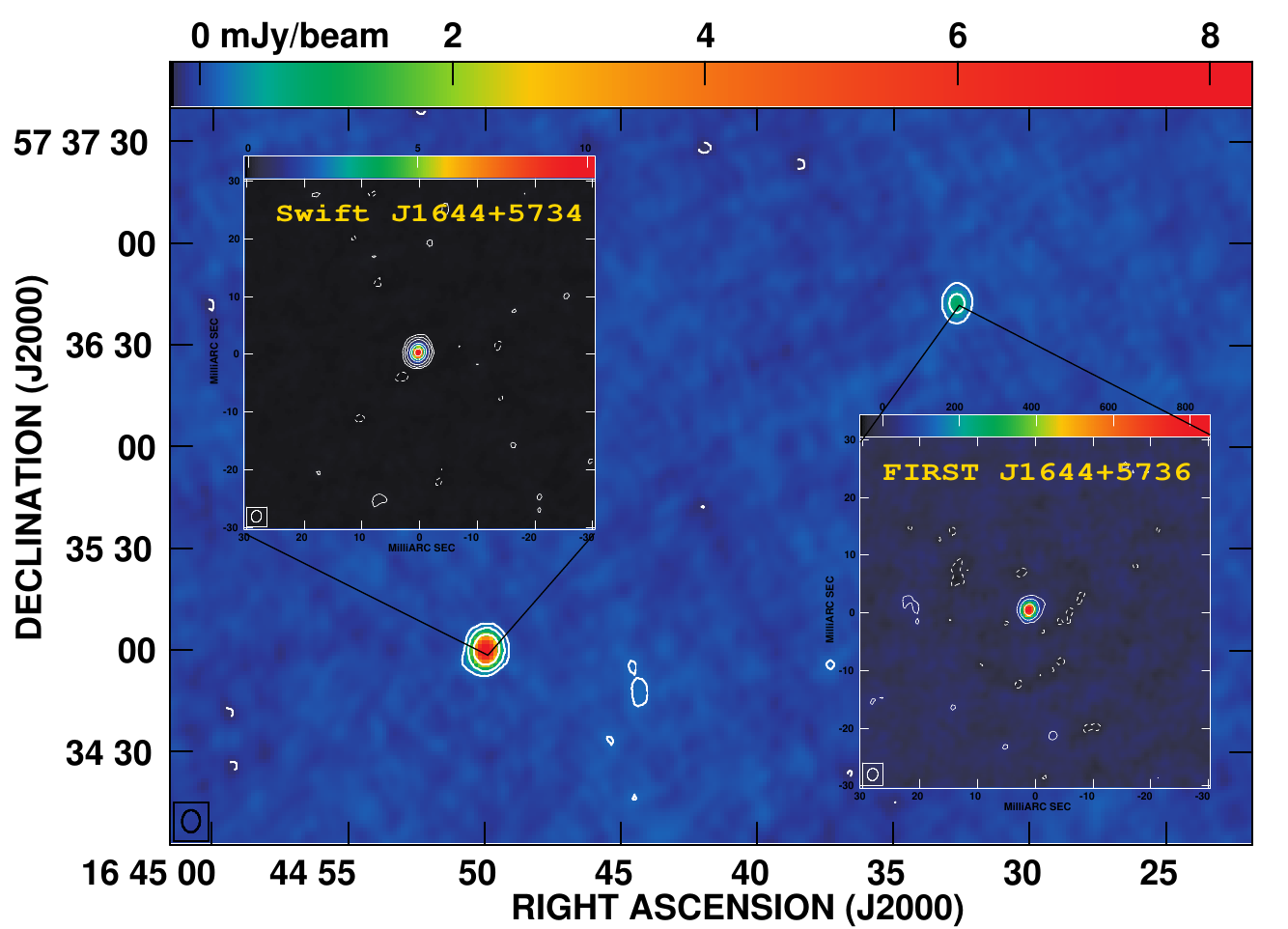} 
 \caption{EVN images of the radio counterpart to Swift J1644+5734 and the in-beam reference source 
 (insets), with a WSRT image as background. From \cite{Y16}, Fig.~1.}
   \label{fig1}
\end{center}
\end{figure}

\section{The first jetted TDE}

The transient Swift J1644+5734 was extremely luminous from radio to the $\gamma$ rays, that has been widely
interpreted as a non-thermal TDE with a relativistic jet (\cite[Levan et al. 2011]{L11}; \cite[Bloom et al. 2011]{BLO11};
\cite[Burrows et al. 2011]{BUR11}; \cite[Zauderer et al. 2011]{ZAU11}). The variability properties in the different bands
indicated that the X-ray and the radio emission were most likely produced in different regions, the latter related to a
decelerating external shock in an initially ultrarelativistic outflow (\cite[Giannios \& Metzger 2011]{GM11}). Similar
models have been invoked to explain multi-band properties of $\gamma$-ray afterglows (e.g. 
\cite[Zauderer et al. 2013]{ZAU13}). The VLBI technique has the power to directly reveal relativistic ejecta by   
for example resolving them.  In case of J1644+5734  \cite{BER12} predicted that marginally resolving the source
may be possible at high frequencies with current VLBI instrumentation; at 22~GHz and above however
the emission is expected to fade quickly. Another way, often successfully applied in quasars up to even 
very high redshifts (see \cite[Frey et al. 2015]{FRE15}), is to detect the proper motion of relativistic ejecta. 
We followed this alternative route.

\section{European VLBI Network (EVN) observations}

We observed Swift J1644+5734 at six epochs between 12 April 2011 and 9 March 2015 with the EVN at 5~GHz. 
Detection of targets at the few-mJy level requires frequent visits to a bright, compact calibrator, a technique called
phase-referencing (see \cite[Reid \& Honma 2014]{RH14}). This technique also provides the relative position between 
the target and the calibrator. For this purpose we used one of the International Celestial Reference Frame (ICRF) 
defining sources, J1638+5720 as primary calibrator. The precision of the derived relative position however depends 
strongly on the separation between these sources (in our case 55 arcminutes). The larger the angular distance, the 
larger the systematic errors are. Therefore our strategy was to identify a more nearby compact source in our very 
first observation, that could be used as a reference point in the following epochs. We did find a candidate in the 
FIRST catalogue that was unresolved on milliarcsecond scales, FIRST J1644+5736, with a separation of 2.9 arcminutes.  
Swift J1644+5734 and FIRST J1644+5736 were within the primary beam of the smaller telescopes, and only the large 
dishes had to regularly nod between these sources (see Fig.\,\ref{fig1} for the EVN images).

\section{Constraints from VLBI}

\begin{figure}[t]
\begin{center}
  \includegraphics[width=4.1in]{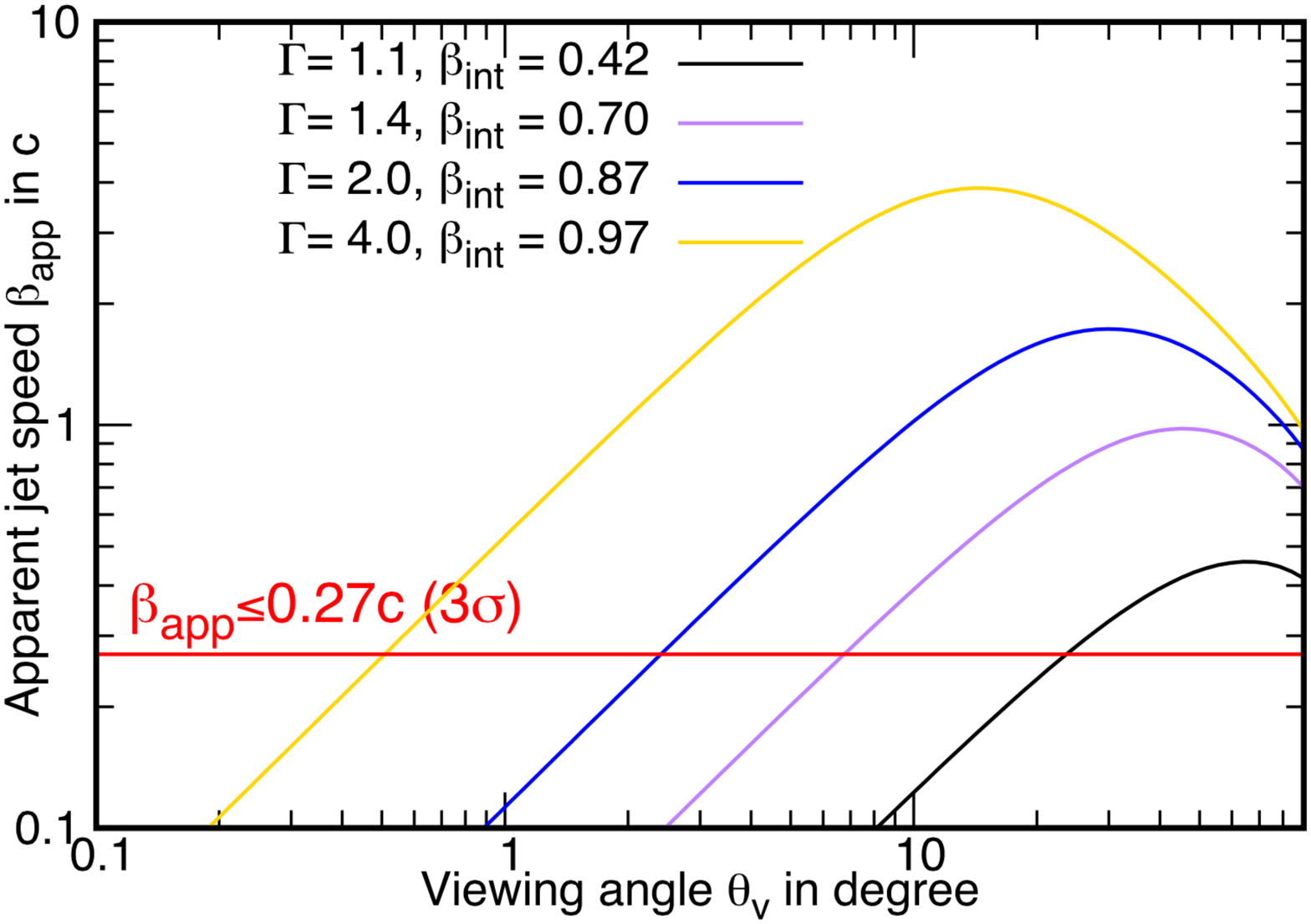} 
\vspace*{-0.2 cm}
 \caption{Constraints on Swift J1644+5734 average intrinsic jet velocity $\beta_{\rm int}$ and 
 Lorentz factor for various viewing angles, from the proper motion upper limit (horizontal line).
 From \cite{Y16}, Fig.~3.}
   \label{fig2}
\end{center}
\end{figure}

Using the technique described above, we achieved very high precision relative astrometry between our in-beam 
reference source and the radio \lq\lq afterglow'' of Swift J1644+5734. The standard deviation of our five relative 
positions spanning three years 
were 13~$\mu$as in Right Ascension and 11 $\mu$as in Declination, just a tiny fraction (1/160th) of the synthesized 
beam. We see no detectable proper motion in Swift J1644+5734 (relative to FIRST J1644+5736). The 3-$\sigma$ 
upper limit to the average (over 3 years) apparent jet velocity is $\beta_{\rm app}<0.27$. The implied limit on the 
intrinsic jet speed as a function of viewing angle is shown on Fig.\,\ref{fig2}. 

If the radio emission is produced by a forward shock analogous to gamma-ray burst jets with a Lorentz-factor 
$\Gamma=2$ (blue line in Fig.\,\ref{fig2}) or higher (cf. \cite[Zauderer et al. 2013]{ZAU13}), the implied viewing angle 
is less than 3 degrees. A lower average Lorentz factor (e.g. due to  a dense interstellar medium) would allow larger 
viewing angles. In this case, the observations would indicate that significant deceleration took place within 
about three years after relativistic ejecta were triggered in Swift J1644+5734, and this may be expected in other 
non-thermal (jet-forming) TDEs as well. Ultra-precise VLBI astrometry commenced at a closer epoch to the trigger 
event may give stronger constraints for other TDEs in the future, possibly for systems that are less-beamed 
(i.e. have a larger viewing angle, but not completely de-boosted). 

Alternatively, the radio jet in Swift J1644+5734 might have been mildly relativistic ($\beta_{\rm int}<0.3$) from the 
beginning, in accordance with what is expected for purely radiatively driven jets (\cite[Sadowski \& Narayan 2015]{SN15}). 
While \cite{K16} detected relativistic reverberation in the accretion flow of the system, and argued for a mildly relativistic
wind as the source of X-rays. These models however do not naturally explain the broad-band properties of 
Swift J1644+5734, that imply a jet with an initially high Lorentz factor.

Relativistic-jet TDEs remain in the forefront of transient research with the EVN, and the first result claiming a resolved 
jet in ASASSN-14li has just been published by \cite{RC16}. In the future, we intend to launch a SKA-VLBI 
(\cite[Paragi et al. 2015]{PAR15}) follow-up program for the most promising candidates found by the SKA.

\end{document}